\documentclass[12pt]{article}
\usepackage{amssymb,amsmath,epsfig}

\begin{document}

\title{\bf Interacting Modified Holographic Dark
Energy in Kaluza-Klein Universe}
\author{M. Sharif \thanks {msharif.math@pu.edu.pk} and Abdul Jawad\thanks {jawadab181@yahoo.com}\\
Department of Mathematics, University of the Punjab,\\
Quaid-e-Azam Campus, Lahore-54590, Pakistan.}

\date{}

\maketitle
\begin{abstract}
The interaction of modified holographic dark energy and dark matter
with varying $G$ in flat Kaluza Klein universe is considered.
Further, we take infrared cutoff scale $L$ as future event horizon.
In this scenario, equations of state parameter as well as evolution
are explored. We also check the validity of the generalized second
law of thermodynamics. It is interesting to mention here that our
results show consistency with the present observations.
\end{abstract}
\textbf{Keywords:} Kaluza-Klein cosmology; Modified holographic
dark energy; Dark matter; Generalized second law of thermodynamics.\\
\textbf{PACS:} 04.50.Cd; 95.36.+d; 95.35.+x; 98.80.-k

\section{Introduction}

It is believed that our universe has entered in an accelerated
expansion phase and experienced a large negative pressure (Riess
et al. 1998; Perlmutter et al. 1999; Fedeli et al. 2009; Caldwell
and Doran 2004). The force responsible for driving the universe
apart is some cosmological antigravity substance known as the
mysterious dark energy (DE). Although $75$ percent of the
mass-energy content of our universe contains this type of DE but
its nature is still speculative. One usually characterizes the DE
phenomena with equation of state (EoS) parameter $\omega$ (the
ratio of pressure to energy density): $\omega$ lying in the range
(-1,-1/3) describes quintessence DE era, $\omega=-1$ represents DE
due to cosmological constant consistent with current observations
(Davis et al. 2007), while it may be $\omega<-1$, the phantom case
(which can lead to a future unavoidable singularity of spacetime).

The simplest candidate of DE is the cosmological constant but it
faces some serious problems (Weinberg 1989, 2000) like 'fine
tuning problem' (unusual small value) and 'coincidence problem'
(why DE and DM are of the same order today even though universe is
expanding$?$). The dynamical DE scenario is an alternative way to
alleviate or even solve these problems. So far, a plethora of
dynamical DE models has been proposed by physicist which are
classified into two categories: scalar field models containing
quintessence (Ratra and Peebles 1988), phantom (Caldwell 2002;
Carroll et al. 2003), k-essence (Armendariz-Picon et al. 1999,
2001; Chiba, et al. 2000), tachyon (Padmanabhan 2002; Bagla et al.
2003), quintom (Feng, et al. 2005) and the interacting DE models
(interaction of DE with DM) including family of Chaplygin gas
(Kamenshchik et al. 2001; Bento et al. 2002; Zhang et al. 2006),
braneworld (Deffayet et al. 2002; Sahni and Shtanov 2003),
holographic DE (HDE) (Hsu 2004; Li 2004), new agegraphic DE models
(Cai 2007) etc. These possibilities reflect the indisputable fact
that the true nature and origin of DE has not been convincingly
explained yet.

In the aforementioned candidates of DE, holographic dark energy is
one of the marvelous attempts to examine the nature of DE in the
framework of quantum gravity. It is based on holographic principle
(Susskind 1995) which states that \textit{all the information
relevant to a physical system inside a spatial region can be
observed on its boundary instead of its volume}. The main feature
of this DE is that it links DE density to the cosmic horizon (a
global property of the universe). Cohen et al. (1999) argued that
the quantum zero-point energy of a system with size $L$ (or
infrared (IR) cutoff) should not exceed the mass of a black hole
with the same size, i.e., $L^{3}\rho_{v}\leq L M^{2}_{pl}$ (where
$\rho_{v}$ indicates the quantum zero-point energy density and
$M_{pl}=(8\pi G)^{-1/2}$ is the reduced Planck mass). Its original
form is defined as (Hsu 2004)
\begin{equation*}
\rho_{\Lambda}=\frac{3m^{2}}{8\pi GL^{2}},
\end{equation*}
where $m$ is constant and G is the gravitational constant which is
taken as a function of time.

The variation of G with time is the natural consequence of Dirac'
"Large Number Hypothesis" (LNH) (Dirac 1938). Following this
hypothesis, many efforts have been made through different aspects
to obtain physical results about the universe. The idea of
Brans-Dicke (BD) theory (Brans and Dicke 1961) and its
generalization to other forms of scalar-tensor theories (Bergmann
1968; Wagoner 1970; Nordtvedt 1970) arose from variable-G theories
in which gravitational constant is replaced by a scalar field
coupling to gravity through a new parameter. Further, it was shown
through different observations in favor of hypothesis that G could
be varied as a function of time or equivalently of the scale
factor (Umezu et al. 2005; Nesseris and Perivolaropoulos 2006;
Biesiada and Malec 2004). Recent work (Setare 2006a, 2006b; Jamil
et al. 2009) shows keen interest to take interacting HDE with or
without varying $G$ in order to explain the current status of the
universe. Some people (Setare 2006c; Setare and Shafei 2006;
Setare and Vagenas 2008) have also explored the generalized second
law of thermodynamics (GSLT) with interacting HDE. Also, this work
was extended in different modified gravities like f(R) theory
(Setare 2008), BD theory (Setare and Jamil 2010a), Gauss-Bonnet
theory (Setare and Jamil 2010b), Horava-Lifshitz theory (Setare
and Jamil 2010c) and Kaluza-Klein theory (Sharif and Khanum 2011).

Kaluza-Klein (KK) theory is an extra dimensional theory which grew
out of an attempt to couple gravity and electromagnetism (Kaluza
1921, Klein 1926). Further development classified this theory into
two versions such as compact (fifth dimension is length like and
it should be very small) and non-compact forms (fifth dimension is
mass like) (Wesson 1984, 1999; Bellini 2003). The non-compact KK
theory is the consequence of well-known Campbell's theorem in
which one cannot introduce any matter into five dimensional ($5D$)
manifold by hand because matter appears in four dimensions induced
by the $5D$ vacuum theory (Wesson 1984, 1999; Bellini 2003).
However, original KK idea became the base of other extra
dimensional theories in different aspects like string theory
(Polchinski 1998), brane models (Clifford 2003) and super gravity
(Wess and Bagger 1992). The version of HDE in extra dimension is
known as modified holographic dark energy (MHDE) (Gong and Li
2010) which can be derived from mass of black hole in $N+1$
dimensional spacetime (Myers 1987). Some authors (Gong and Li
2010, Liu et al. 2010) found interesting results about the
evolution of the universe by considering MHDE model. Recently,
Sharif and Khanum (2011) have discussed the interacting MHDE with
DM in KK cosmology by taking IR cutoff as Hubble horizon and also
explored that GSLT holds generally without any assumption in this
scenario.

The selection of IR cutoff $L$ is a crucial problem inherited in
the original version of HDE model which could be taken as either
Hubble horizon $H^{-1}$, or particle horizon or future event
horizon. Hsu (2004) and Li (2004) showed that if $L$ is taken as
either Hubble horizon $H^{-1}$, or particle horizon then the DE
model is not compatible with the current observational data.
However, it was suggested by Li (2004) that future event horizon
is the best choice for $L$.

This work is an extension of Sharif and Khanum (2011) with IR
cutoff as future event horizon and also varying $G$. Equations of
state parameter as well as evolution are formulated. We also
investigate the validity of GSLT in this scenario. The paper is
organized as follows: Section \textbf{2} is devoted to the
formulation of interacting MHDE with DM and its discussion. In
section \textbf{3}, we study the validity of GSLT for interacting
models. In the last section, we summarize our results.

\section{Interacting Modified Holographic Dark
Energy with Varying G}

In this section, we discuss the interaction of MHDE with dark
matter in Kaluza-Klein cosmology by taking $G$ as a function of
cosmic time $t$. We evaluate EoS parameter and evolution of MHDE.
The non-flat KK universe (Ozel et al. 2010) is given by
\begin{equation}\label{1}
ds^{2}=dt^{2}-a^{2}(t)[\frac{dr^{2}}{1-kr^{2}}
+r^{2}(d\theta^{2}+\sin\theta^{2}d\phi^{2})+(1-kr^{2})d\psi^{2}],
\end{equation}
where a(t) is the dimensionless scale factor measuring the expansion
of the universe, $k=-1,0,1$ is the curvature parameter for the open,
flat and closed universe respectively. Equations of motion
corresponding to (\ref{1}) for flat KK universe are
\begin{eqnarray}\label{2}
\frac{\dot{a}^{2}}{a^{2}}&=&\frac{8\pi G(t)}{6}\rho,\\\label{3}
\frac{\ddot{a}}{a}&=&-\frac{8\pi G(t)}{6}(\rho+2p).
\end{eqnarray}
Here dot represents differentiation with respect to cosmic time $t$
and $p=p_{\Lambda},~\rho=\rho_{\Lambda}+\rho_m$ are the pressure and
density respectively (where subscripts $\Lambda$ and $m$ denote DE
and DM respectively). Also, we define $H\equiv\dot{a}/{a}$ as the
Hubble parameter which estimates the expansion rate of the universe.

In this scenario, the fractional energy densities are
\begin{eqnarray}\label{4}
\Omega_{m}=\frac{8\pi G\rho_{m}}{6H^{2}},\quad
\Omega_{\Lambda}=\frac{8\pi G\rho_{\Lambda}}{6H^{2}}
\end{eqnarray}
so that Eq.(\ref{2}) takes the form
\begin{equation}\label{5}
\Omega_{m}+\Omega_{\Lambda}=1.
\end{equation}
The energy conservation for the KK universe becomes
\begin{equation*}
\dot{\rho}+4H(\rho+p)=0.
\end{equation*}
Since we consider mixture of DE and DM (dust like), the above
conservation equation does not hold globally. It should be
decomposed into two non-conserving equations for DM and MHDE
respectively, i.e.,
\begin{eqnarray}\label{6}
\dot{\rho}_{m}+4H\rho_{m}=\Upsilon_{4},\quad
\dot{\rho}_{\Lambda}+4H(\rho_{\Lambda}+p_{\Lambda})=-\Upsilon_{4},
\end{eqnarray}
where $\Upsilon_4$ is the energy exchange term. The available
three forms of energy exchange term in literature (Sun 2010) are
given as
\begin{eqnarray*}
\Upsilon_{1}=3dH\rho_{\Lambda},\quad
\Upsilon_{2}=3dH(\rho_{\Lambda}+\rho_{m}),\quad
\Upsilon_{3}=3dH\rho_{m},
\end{eqnarray*}
where $d$ is a coupling constant which has a crucial role in the
interpretation of interaction processes, i.e., either DM decays
into DE or DE decays into DM depending on the sign of $d$ .

The positive value of $d$ stands for decay from DE to DM and
negativity of $d$ exhibits its reverse process. Moreover, the
negative value of the coupling constant $d$ would lead $\rho_{m}$
to negative for far future when $\Upsilon=\Upsilon_{1}$ and
$\Upsilon=\Upsilon_{2}$. For $\Upsilon=\Upsilon_{3}$ with negative
$d$, it does not suffer such type of difficulties. But Pavon and
Wang (2009) pointed out that GSLT favors the criteria of decays
from DE to DM. Sun (2010) proposed a new form of energy exchange
term
\begin{equation}\label{7}
\Upsilon_{4}=3dH(\rho_{\Lambda}-\rho_{m}),
\end{equation}
which is the best suitable model with the current observations.
The value of $d$ is assumed to be positive due to the reasons
mentioned above. The interesting feature of this interacting model
is that it shows consistency with GSLT according to the arguments
(Pavon and Wang 2009) both for early as well as late time.

In the following, we develop the modified version of holographic
dark energy in KK cosmology. For this purpose, we consider a
relationship of mass and radius of the Schwarzschild black hole in
$N+1$ dimensions as (Myers 1987)
\begin{equation*}
M=\frac{(N-1)A_{N-1}R^{N-2}_{H}}{16\pi G},
\end{equation*}
where $A_{N-1}$ represents the area of unit $N$-sphere, $R_{H}$
stands for the horizon scale of black hole. Also, G ($N+1$
dimensional gravitational constant) is associated with $M_{N+1}$
($N+1$ dimensional Planck mass) and the usual Planck mass $M_{pl}$
in $4$-dimensional spacetime through
\begin{equation*}
8\pi G=M^{-(N-1)}_{N+1}=\frac{V_{N-3}}{M^{2}_{pl}},
\end{equation*}
where $V_{N-3}$ is the volume of the extra-dimensional space. Thus
the mass becomes
\begin{equation*}
M=\frac{(N-1)A_{N-1}R^{N-2}_{H}M^{2}_{pl}}{2V_{N-3}}.
\end{equation*}
The mass related to the MHDE is given as (Myers 1987)
\begin{equation*}
\rho_{\Lambda}=\frac{m^{2}(N-1)A_{N-1}L^{N-5}M^{2}_{pl}}{2V_{N-3}}.
\end{equation*}
For KK cosmology, we substitute $N=4$ and the value of $A_{3}$ in
the above expression, it follows that
\begin{equation}\label{8}
\rho_{\Lambda}=\frac{3m^{2}\pi^{2}L^{2}}{8\pi G}.
\end{equation}

The future event horizon $L$ is defined as (Li 2004)
\begin{equation*}
L=a(t)\int^{\infty}_{a}{\frac{d\acute{a}}{H\acute{a}^{2}}}=
a(t)\int^{\infty}_{a}{\frac{d\acute{t}}{a(\acute{t})}}.
\end{equation*}
Its derivative with respect to time yields
\begin{equation}\label{9}
\dot{L}=HL-1.
\end{equation}
For the de Sitter spacetime, the future event horizon becomes
$H^{-1}$ and $\dot{L}=0$. Thus the future event horizon and
apparent horizon coincide with each other for the spatially flat
de Sitter universe and there is only one cosmological horizon.

Using Eqs.(\ref{4}) and (\ref{9}), we obtain time derivative of
energy density
\begin{equation}\label{10}
\dot{\rho}_{\Lambda}=-\rho_{\Lambda}H(-2+\frac{2m\pi}{H^{2}\sqrt{2\Omega_{\Lambda}}}
+\Delta_{G}),
\end{equation}
where $\Delta_{G}\equiv\frac{G'}{G},~\dot{G}=G'H$, here prime
denotes derivative with respect to $\ln a(t)$. Using EoS of DE,
$p_{\Lambda}=\omega_{\Lambda}\rho_{\Lambda}$, in Eq.(\ref{6}), we
have
\begin{eqnarray}\label{11}
\dot{\rho}_{m}+4H(1+\omega^{eff}_{m})\rho_{m}&=&0,\\\label{12}
\dot{\rho}_{\Lambda}+4H(1+\omega^{eff}_{\Lambda})\rho_{\Lambda}&=&0,
\end{eqnarray}
where
\begin{equation}\label{13}
\omega^{eff}_{m}=-\frac{\Upsilon_{4}}{4H\rho_{m}},\quad
\omega^{eff}_{\Lambda}=\omega_{\Lambda}+\frac{\Upsilon_{4}}{4H\rho_{\Lambda}}
\end{equation}
are effective equations of state for DM and MHDE respectively.
Inserting the values of $\dot{\rho}_{\Lambda}$ and $\Upsilon_{4}$ in
Eq.(\ref{12}) and using Eq.(\ref{4}), we obtain
\begin{equation}\label{14}
\omega_{\Lambda}=-\frac{3}{2}+\frac{m\pi}{2H^{2}\sqrt{2\Omega_{\Lambda}}}
+\frac{1}{4}\Delta_{G}-\frac{3d}{4}(2-\frac{1}{\Omega_{\Lambda}}).
\end{equation}
This is the EoS for the MHDE which depends upon Hubble parameter
$H$, dimensionless DE density $\Omega_{\Lambda},~\Delta_{G}$ and
also two positive constant parameters $m,~d$.

According to the present status of the universe, it was observed
that $H_{0}=1$ (here $0$ denotes the present value) (Paul et al.
2009), $\Omega_{\Lambda}=0.73$ and the value of $\Delta_{G}$ lies
in the range $0<\Delta_{G}\leq0.07$ (Setare 2006a; Jamil et al.
2009). Also, Sun (2010) investigated that the value of interacting
parameter $d=0.001$ is consistent with the analysis of (Pavon and
Wang 2009) which shows consistency with the known properties of
DE. However, the value of constant parameter $m$ is still
ambiguous. The best value of $m$ obtained from observational type
Ia supernovae is $0.21$ (Huang and Gong 2004) while it leads to
$0.61$ from the X-ray gas mass fraction of galaxy clusters (Chang
et al. 2006). Both of these values of $m$ are taken within
$1-\sigma$ error range. Also, it was consensus on the value of
$m=0.91$ taken from combining the observational data of type Ia
supernovae, cosmic microwave background radiation and large scale
structure within $1-\sigma$ error range (Zhang and Wu 2005).
Later, its value $m=0.73$ was observed through combining the data
of type Ia supernovae, X-ray gas and Baryon Acoustic Oscillation
(Wu et al. 2008; Ma and Gong 2009).

Using these values of the parameters, especially of $\Delta_{G}$ and
$m$, we evaluate the value of EoS parameter which remains less than
$-1/3$, i.e., $\omega_{\Lambda}\leq -1/3$ for $m=0.61, 0.73, 0.91$
(which follows the quintessence DE region). For $m=0.21$, we have
phantom DE era, i.e., $\omega_{\Lambda}< -1$. These ranges of
$\omega_{\Lambda}$ describe the accelerating expansion of the
universe through DE as a driving force. Thus the MHDE shows the
transition from quintessence DE era to phantom DE era under certain
assumptions.

The fractional energy density of MHDE is
\begin{equation}\label{15}
\Omega_{\Lambda}=\frac{8\pi G\rho_{\Lambda}}{6H^{2}}.
\end{equation}
Also, we have
\begin{equation}\label{16}
\frac{\dot{2H}}{H^{2}}=-4(1+\omega_{\Lambda}\Omega_{\Lambda})+\Delta_{G}.
\end{equation}
Taking derivative of Eq.(\ref{15}) with respect to cosmic time and
using Eq.(\ref{16}), it follows that
\begin{equation}\label{17}
\dot{\Omega}_{\Lambda}={\Omega}_{\Lambda}H(6+\omega_{\Lambda}\Omega_{\Lambda}-\Delta_{G}
-\frac{m\pi}{2H^{2}\sqrt{2\Omega_{\Lambda}}})
\end{equation}
which can also be written as
\begin{equation}\label{18}
\Omega'_{\Lambda}={\Omega}_{\Lambda}(6+\omega_{\Lambda}\Omega_{\Lambda}-\Delta_{G}
-\frac{m\pi}{2H^{2}\sqrt{2\Omega_{\Lambda}}}).
\end{equation}
This is the evolution equation for the MHDE which is an increasing
function and represents the gradual increment of DE in the universe.

\section{Generalized Second Law of Thermodynamics}

According to this law, the sum of entropy of matter inside horizon
and entropy of the event horizon cannot decrease with time
(Izquierdo and Pav$\acute{o}$n 2006). The thermodynamics of
cosmological scenario requires to consider the universe as a
thermodynamical system and it is related with the thermodynamics
of black hole. The analysis of Hawking (1975) and Gibbons and
Hawking (1977) about black hole implies that temperature of black
hole horizon is inversely proportional to its mass. Bekenstein
(1973) argued that the entropy of black hole would decrease with
time and violate the second law of thermodynamics due to
evaporation (in fact entropy of black hole horizon and horizon
area are directly proportional). Due to this phenomenon, the
entropy of the background universe increases with time. He
concluded that the sum of black hole entropy and the background
entropy must be an increasing quantity with respect to time.

In the following, we discuss the validity of the GSLT of
interacting MHDE with DM with varying G in the KK universe
enclosed by future event horizon. Gibb's equation (Izquierdo and
Pav$\acute{o}$n 2006) relates the entropy of the universe
including DE and DM inside the horizon with its pressure and
internal energy (which is also known as first law of
thermodynamics), i.e.,
\begin{equation*}
TdS=pdV+dE,
\end{equation*}
where $T,~S,~E$ and $p$ are temperature, entropy, internal energy
and pressure of the system respectively and $V=\frac{\pi^{2}
L^{4}}{2}$ is the volume of the system. Splitting the above
expression into MHDE and DM as
\begin{equation}\label{19}
TdS_{\Lambda}=p_{\Lambda}dV +dE_{\Lambda},\quad TdS_{m}=p_{m}dV
+dE_{m}.
\end{equation}
Since the number of particles inside the horizon is not conserved,
so we assume chemical potential to be zero (Izquierdo and
Pav$\acute{o}$n 2006). The temperature and entropy of horizon in
KK universe become (Cai and Kim 2005)
\begin{equation*}
T=\frac{1}{2\pi L},\quad S_{H}=\frac{\pi^{2}L^{3}}{2G}.
\end{equation*}
The rate of change of entropy of horizon is
\begin{equation}\label{20}
\dot{S_{H}}=\frac{3\pi^{2}L^{2}\dot{L}}{2G}-\frac{\pi^{2}L^{3}\dot{G}}{2G^2}.
\end{equation}

Thermodynamical and cosmological quantities are related as
\begin{eqnarray}\label{21}
p_{\Lambda}=\omega^{eff}_{\Lambda}\rho_{\Lambda},\quad
p_{m}=\omega^{eff}_{m}\rho_{m},\quad E_{\Lambda}=\frac{\pi^{2}
L^{4}\rho_{\Lambda}}{2},\quad E_{m}=\frac{\pi^{2}
L^{4}\rho_{m}}{2}.
\end{eqnarray}
Taking the derivative of Eq.(\ref{19}) with respect to time, we have
\begin{equation}\label{22}
\dot{S_{\Lambda}}=\frac{p_{\Lambda}\dot{V}+\dot{E_{\Lambda}}}{T},\quad
\dot{S_{m}}=\frac{p_{m}\dot{V}+\dot{E_{m}}}{T}.
\end{equation}
Equations (\ref{20})-(\ref{22}) lead to
\begin{eqnarray}\nonumber
\dot{S}_{total}&=&
4\pi^{3}L^{4}[(1+\omega^{eff}_{\Lambda})\rho_{\Lambda}
+(1+\omega^{eff}_{m})\rho_{m}](\dot{L}-LH)+\frac{3\pi^{2}L^{2}\dot{L}}{2G}\\
&-&\frac{\pi^{2}L^{3}\dot{G}}{2G^2}\label{23},
\end{eqnarray}
where $S_{total}$ is the sum of three entropies. Also, we know that
\begin{equation}\label{24}
(1+\omega^{eff}_{\Lambda})\rho_{\Lambda}+(1+\omega^{eff}_{m})\rho_{m}
=(1+\omega_{\Lambda})\rho_{\Lambda}+\rho_{m}
\end{equation}
and
\begin{equation}\label{25}
\frac{\rho_{m}}{\rho_{\Lambda}}=-1+\frac{1}{\Omega_{\Lambda}}.
\end{equation}
Finally, we obtain the following expression by using Eqs.(\ref{9}),
(\ref{24}) and (\ref{25}) in (\ref{23})
\begin{equation}\label{26}
\dot{S}_{total}=\frac{3m^2\pi^4L^6}{8G}
\left(-\omega_{\Lambda}-\frac{1}{\Omega_{\Lambda}}
-\frac{m^2\pi^2}{4H^{4}\Omega^{2}_{\Lambda}}
+\frac{4m\pi}{(2\Omega_{\Lambda})^{\frac{3}{2}}H^2}(1-\Delta_{G})\right).
\end{equation}
If we set the values of $\Omega_{\Lambda},~\Delta_{G}$ and
$\omega_{\Lambda}$ corresponding to all physical acceptable values
of $m$ discussed in the previous section, we find
$\dot{S}_{total}\geq0$. Thus the GSLT holds for KK universe
containing interaction of MHDE and DM with varying $G$ enclosed by
the cosmological future event horizon.

\section{Summary}

We have explored interaction of MHDE with DM and generalized
second law of thermodynamics in KK universe. Setare (2006b)
obtained an expression of EoS for holographic dark energy by
considering bulk brane Interaction and found that it may cross
phantom divide $\omega=-1$. The same author (2009) pointed out
that holographic Chaplygin gas model, with IR cutoff as future
event horizon, behaves alike phantom fluid and cross the phantom
divide in a Dvali-Gabadaze Porrati (DGP) braneworld framework. Liu
et al. (2010) and Bandyopadhyay (2011) found that EoS of the MHDE
can cross the phantom divide from quintessence region to phantom
region during the evolution by choosing future event horizon as
the IR cutoff in a DGP braneworld scenario. It was showed (Setare
and Shafei 2006; Setare and Vagenas 2008) that GSLT is respected
under the special range of physical parameters for HDE with
horizon's radius $L$ measured from the sphere of horizon. Also, it
was found (Dutta and Chakraborty 2010, 2011) for DGP braneworld
interacting HDE and CDM, GSLT holds for the universe bounded by
apparent horizon or future event horizon.

In this paper, we have taken interacting MHDE and DM with varying
$G$ in the platform of KK cosmology. Further, we assume that IR
cutoff $L$ in the MHDE as a future event horizon. We have found
the EoS parameter $\omega_{\Lambda}$ of MHDE depending on
different parameters by using the interacting model suggested by
Sun (2010). In particular, the constant parameter $m$ plays the
crucial role in evaluating $\omega_{\Lambda}$. It is found that
$\omega_{\Lambda}$ remains in the quintessence DE dominated era
for $m=0.61,~0.73,~0.91$ and it enters into the phantom DE
dominated era for $m=0.21$. This means that the universe shows
transition from quintessence DE to phantom DE era in the scenario
of MHDE. With the expansion of the universe, it is determined that
the varying $G$ with time is an increasing function of time.
Although the variations are negligibly small, so it does not
affect on our major results.

We have also investigated the validity of the generalized second law
of thermodynamics in this scenario. It turns out that GSLT holds for
the specific choice of physical parameters. It is interesting to
mention here that our results about evolution of the universe are
consistent with the current observations (Liu et al. (2010), Setare
(2009), Bandyopadhyay (2011) and also GSLT holds (Dutta and
Chakraborty 2010, 2011) when EoS of MHDE can cross the phantom
divide.

\end{document}